\documentstyle[11pt,a4,palatino]{article}
\input psfig.tex
\begin{document}
\input psfig.tex
\begin{center}
{\large\bf One Parameter Solution of Spherically Symmetric Accretion 
in Various Pseudo-Schwarzschild Potentials}\\[1.5cm]
\end{center}
\begin{center}
{\bf Aveek Sarkar} \\[0.25cm]
{\it National Centre For Radio Astrophysics, TIFR, 
Post Bag 3 Ganeshkhind, \\
Pune 411 007, India\\ 
Email: sarkar@ncra.tifr.res.in}\\ [0.25cm]
{\bf Tapas K. Das}\\[0.25cm]
{\it Inter University Centre For Astronomy And Astrophysics,
Post Bag 4 Ganeshkhind, Pune 411 007, India\\
Email:  tapas@iucaa.ernet.in}\\[1cm]
\end{center}
\begin{center}
{\bf Abstract} 
\footnote{Send any correspondense/ offprint requests to Tapas K. Das
({\it tapas@iucaa.ernet.in})}\\[0.5cm]
\end{center}
\noindent
In this paper we have solved the hydrodynamic equations governing the
spherically symmetric isothermal accretion (wind) onto (away from) compact
objects using various pseudo-Schwarzschild potentials.These solutions are
essentially one parameter solutions in a sense that all relevant dynamical
as well as thermodynamic quantities for such a flow could be obtained
(with the assumption of a one-temperature fluid)
if {\it only one} flow parameter (temperature of the flow $T$) is given.
Also we have investigated the transonic behaviour of such a flow
and showed that for a given $T$, transitions from subsonic to the
supersonic branch of accretion (wind) takes place at different locations
depending on the potentials used to study the flow and we have identified
these transition zones for flows in various such potentials.\\[0.5cm]
\hrule
\noindent
{\bf Published in the International Journal of Modern Physics D, 2002,
Volume 11, Issue 03, pp. 427-432.} 
\hrule
\vskip 0.5cm
\section {Introduction}
A number of modified Newtonian potentials of various forms are available 
in literature
which accurately approximate some general relativistic effects important
to study accretion discs around a Schwarzschild black hole.
Such potentials may be called `pseudo-Schwarzschild' potentials because they 
nicely
mimic the space-time around  non-rotating/slowly
rotating compact astrophysical bodies. Recently we have shown that$^1$
(Das \& Sarkar, 2001, hereafter PI) though the potentials
discussed in this paper
 were originally proposed to mimic the
relativistic effects manifested in disc accretion,
it is quite reasonable to use most of these
potentials in studying various dynamical as well as thermodynamic
quantities also
for spherical Bondi-type$^2$ accretion 
of adiabatic fluid onto Schwarzschild black holes.
In PI, the space variation of various dynamical and thermodynamic
quantities had been studied for a two parameter adiabatic 
accretion and wind system i.e., flow parametrized by the specific energy 
and the polytropic constant of the flow. In this paper, we would 
like to continue our investigation of spherical accretion for the same
set of pseudo potentials used in PI but for isothermal accretion (wind) which
may be characterized by a single parameter, i.e., the flow temperature which
can be approximated by the proton temperature for an one-temperature fluid
system.\\
\noindent
Owing to the fact that, close to a spherically accreting Schwarzscild
black hole, the ratio of electron number density to the photon number density
is proportional to $\sqrt{r}$ (where $r$
is the radial distance measured from the central accretor in units of
Schwarzschild radius $r_g=\frac{2GM_{BH}}{C^2}$), a 
lesser number of electrons would available per photon  with decrease of $r$,
and close to the 
accretor, momentum transfer by photons on the accreting fluid (or wind) might
be an efficient process which may keep the flow temperature 
roughly constant (at 
least upto the sonic point) and hence isothermality may not be an entirely 
unjustified assumption$^3$. However, this efficiency for momentum transfer
falls of for large $r$ and far away from the black hole isothermality assumption
may brake down.
\section {Governing equations}
\noindent
We take the following pseudo potentials:
$$
\Phi_{1}=-\frac{1}{2(r-1)}, ~~
\Phi_{2}=-\frac{1}{2r}\left[1-\frac{3}{2r}+12{\left(\frac{1}{2r}\right)}^2
\right]
$$
$$
\Phi_{3}=-1+{\left(1-\frac{1}{r}\right)}^{\frac{1}{2}},~~
\Phi_{4}=\frac{1}{2}ln{\left(1-\frac{1}{r}\right)}
\eqno{(1)}
$$
to study the flow. Whereas $\Phi_1$ and $\Phi_2$ had been proposed by
Paczy\'nski and Wiita$^4$ (1980) and Nowak and Wagoner$^5$
(1991) respectively,
$\Phi_3$ and $\Phi_4$ are suggested by Artemova et. al.$^6$ (1996) (see
[6] and PI for detail discussion about
these potentials). Hereafter, we will denote any $i$th potential
as $\Phi_i$ where $\left\{i=1,2,3,4\right\}$ corresponds to
$\left\{\Phi_{1},\Phi_{2},\Phi_{3},\Phi_{4}\right\}$ respectively.\\
\noindent
For isothermal equation of state, 
$$
P=\frac{R{\rho}T}{\mu}=C_s^2{\rho}
\eqno{(2)}
$$
where $P,{\rho},T$ and $C_s$ are the pressure, density, constant 
temperature and the constant isothermal sound speed of the flow respectively; $R$
and $\mu$ being universal gas constant and the mean molecular weight. 
$C_s$ and $T$ can be related as, 
$$
C_s^2={\Theta}T
\eqno{(3)}
$$
Where $\Theta=\frac{\kappa}{{\mu}m_H}$, a constant, $m_H\sim{m_p}$ being the
mass of the Hydrogen atom. Integrating the radial momentum 
conservation equation\\
$$
\frac{{\partial{u}}}{{\partial{t}}}+u\frac{{\partial{u}}}{{\partial{r}}}+\frac{1
}{\rho}
\frac{{\partial}P}{{\partial}r}+{\Phi_i}^{'}=0
$$
and the equation of continuity\\
$$
\frac{{\partial}{\rho}}{{\partial}t}+\frac{1}{r^2}\frac{{\partial}}{{\partial}r}
\left({\rho}ur^2\right)=0
$$
using eqn. (2) and (3), we obtain two conservation
equations for our flow as (in system of units used in PI):
$$
\frac{u_i^2}{2}+{\Theta}Tln{\rho_i}+\Phi_i={\bf C}
\eqno{(4a)}
$$
$$
{\dot M}=4{\pi}{{\rho}_i}u_ir^2
\eqno{(4b)}
$$
Where ${\bf C}$ is a constant, $u$ and ${\dot M}$ being the dynamical flow
velocity and the mass accretion rate respectively. Subscript $i$
indicates that 
respective quantities are measured for a particular $\Phi_i$.
The space rate of change of dynamical velocity can be written as:
$$
\frac{du_i}{dr}=\frac{{\frac{2{\Theta}T}{r}}-{\Phi_i}^{'}}{u_i-\frac{{\Theta}T}
{u_i}}
\eqno{(4c)}
$$
where $\Phi_i^{'}=\frac{d{\Phi_i}}{dr}$. From eqn. (4b), one can easily
obtain the sonic point conditions as,
$$
u_i=\sqrt{\frac{r_c^i{\Phi_i}^{'}{\Bigg{\vert}}_c}{2}}
=C_s={\Theta}T
\eqno{(4d)}
$$
where ${\Phi_i}^{'}{\Bigg{\vert}}_c$ is the value of $\frac{d{\Phi_i}}{dr}$
at the corresponding sonic point $r_c^i$. $\frac{du_i}{dr}$ at the sonic point
may be obtained by solving the following equation
$$
{{\left(\frac{du}{dr}\right)}^2}_{c_,i}+
0.25\left({{\Phi_i}^{'}}{{\Bigg{\vert}}}_c\right)^2{\Theta}^{-1}T^{-1}+
0.5{{\Phi_i}^{''}}{{\Bigg{\vert}}}_c
=0
\eqno{(4e)}
$$
where ${{\Phi_i}^{''}}{{\Bigg{\vert}}}_c$ is the value of $\frac{d^2\Phi_i}{dr^2}$
at sonic point. Mach number of the flow $M_i$ for a given $T$ 
and $\Phi_i$ can be 
expressed as,
$$
M_i=u_i{\Theta}^{-0.5}T^{-0.5}=
$$
$$
\sqrt{T{\dot M}{\rho_i}^{-2}r^{-2}+2\Phi_i{\Theta}^{-1}T^{-1}+\left(2T^2-1\right)
ln{\rho_i}}
\eqno{(5)}
$$
where $\dot M$ is the mass accretion rate defined in eqn. (4b).
\section{Results}
\noindent
For a given flow
temperature $T$, one can solve eqn. (4d) to find out the sonic 
point $r_c^i$ for any particular $i$th potential $\Phi_i$. In figure 1 we 
show the variation of sonic
points (for $\Phi_{i=1-4}$) with constant flow 
temperature. We observe that for all $\Phi_{i=1-4}$'s, sonic point anticorrelates
with $T$ non-linearly and monotonically. This is obvious because as sound speed
$C_s$ is constant throughout isothermal flow,
 it is only the dynamical velocity $u$
whose change would control the change of Mach number, and as $C_s$ is proportional
to $\sqrt{T}$, the
more will be 
the flow temperature, the more will be $C_s$ and the less will be the
Mach number $M$ ($M=\frac{u}{C_s}$) for a particular radial distance and the 
less will be the location of $r_c$ where $u$ becomes equal to $C_s$, i.e., 
$M$ becomes unity. It is evident from the figure that for a given flow 
temperature (fixed $T$):
$$
r_c^1~>~r_c^4~>~r_c^3~>~r_c^2
\eqno{(6)}
$$
For a fixed $T$, it is now quite straightforward to simultaneously 
solve eqn. (4a-4e) and eqn. 5 to get the integral curves of motion 
for all $\Phi_{i=1-4}$'s to study the transonicity of the flow. In figure 2,
for $T=5\times10^{10}~{^o\!K}$, we plot all the integral curves of motion for
accretion as well as 
for all $\Phi_{i=1-4}$'s described in 
\S 1. For a given flow temperature and for a 
particular distance $r=r_o$, if we define $M_o^i$ to be the 
Mach number attained at $r_o$ for a specific $\Phi_{i}$, it is clear from the
figure that
$$
M_o^1~>~M_o^4~>~M_o^3~>~M_o^2
\eqno{(7)}
$$
while the sequence is just reversed for the wind. As velocity profile 
for the isothermal flow is identical in nature of its Mach number 
profile with a scale shift of $\left({\Theta}T\right)^{-0.5}$, one 
can conclude that $\Phi_1$ produces the steepest velocity 
gradient for accretion and flattest velocity gradient for wind. Combining
eqn. (6) with (7), one can interpret that for isothermal accretion,
Paczy\'nski and Wiita$^4$ (1980) potential $\Phi_1$ would produce the 
maximum spatial rate of change of kinetic energy of the {\it accretion}
for a given $T$ while the same is produced by potential proposed by
Nowak and Wagoner$^5$ (1991), i.e., $\Phi_2$, for {\it wind}. Similarly,
if for any $\Phi_i$, one defines the spatial gradient of
Mach number to be
$\frac{dM^i}{dr}$, which might be regarded as the measure of the 
`transonicity' of the flow (because more will be the value of
$\frac{dM^i}{dr}$, the faster the flow will become supersonic, i.e., the
further will be the sonic point away from the black hole), we find that
for accretion branch, $\left|\frac{dM^i}{dr}\right|$, e.g., the absolute 
value of $\frac{dM^i}{dr}$, increases non-linearly and monotonically as the 
flow moves towards the black hole and sequence for  $\left|\frac{dM^i}{dr}\right|$
follows the same trend shown in eqn. (7).
However, for wind branch, $\left|\frac{dM^i}{dr}\right|$ does {\it not}
monotonically increase as the flow moves away from the accretor rather
$\left|\frac{dM^i}{dr}\right|$ attains a maximum value in the {\it subsonic
part} of the wind and then starts falling non-linearly,
see figure 3. 
The location of the maxima
can be obtained by solving the following non-trivial equation for 
$r_p$ (the location of the peak):
$$
\left(r_p^i\right)^2{{\bf \Psi_p^i}}(r_p,u_p)+
r_p^i{\bf {\Omega_p^i}}(r_p,u_p)-
2{\Theta}T
=0
\eqno{(8)}
$$
Where ${{\bf \Psi^i}}(r,u)$ and ${{\bf \Omega^i}}(r,u)$ are two complicated
functions of $r$ and $u$ defined as:
$$
{{\bf \Psi^i}}(r,u)=\frac{1}{u_i^2}\frac{du_i}{dr}
\left({\frac{u_i^2+{\Theta}T}{u_i^2-{\Theta}T}}\right)\Phi_i
-\Phi_i^{''}
$$
$$
{{\bf \Omega^i}}(r,u)=-\frac{2}{u^2}\frac{du_i}{dr}
\left({\frac{u_i^2+{\Theta}T}{u_i^2-{\Theta}T}}\right){\Theta}T
$$
The superscript $i$ indicates the value obtained for a particular 
$i$th potential. Also we obtain that the location of the peak maintains the
following sequence:
$$
r_p^1~<~r_p^4~<~r_p^3~<~r_p^2
\eqno{(9)}
$$
Thus we successfully analyze various aspects of transonic behaviour of the
spherically symmetric, isothermal accretion characterized by its flow 
temperature $T$.\\ \\
\noindent
{\bf Acknowledgements} \\ \\
\noindent
We are thankful to  Prof. I. D. Novikov and Prof. P. J. Wiita for useful
discussions.

\newpage
\begin{figure}
\vbox{
\vskip -3.0cm
\centerline{
\psfig{file=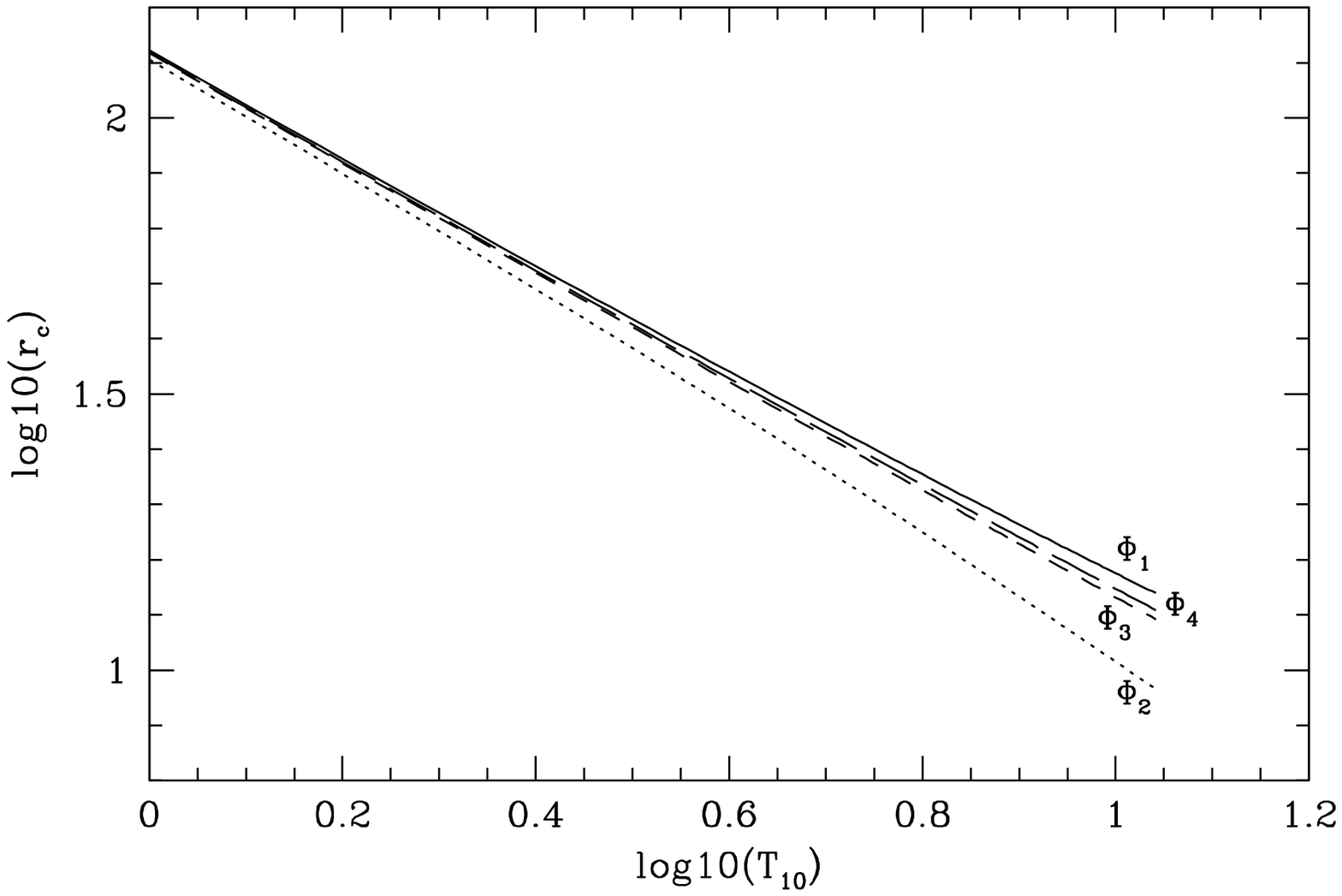,height=20cm,width=20cm}}}
\noindent {{\bf Fig. 1:}\\
{\bf Aveek Sarkar and Tapas K. Das\\
One Parameter Solution of Spherically Symmetric Accretion in Various
Pseudo-Schwarzschild Potentials}}
\end{figure}
\newpage
\begin{figure}
\vbox{
\vskip -4.5cm
\centerline{
\psfig{file=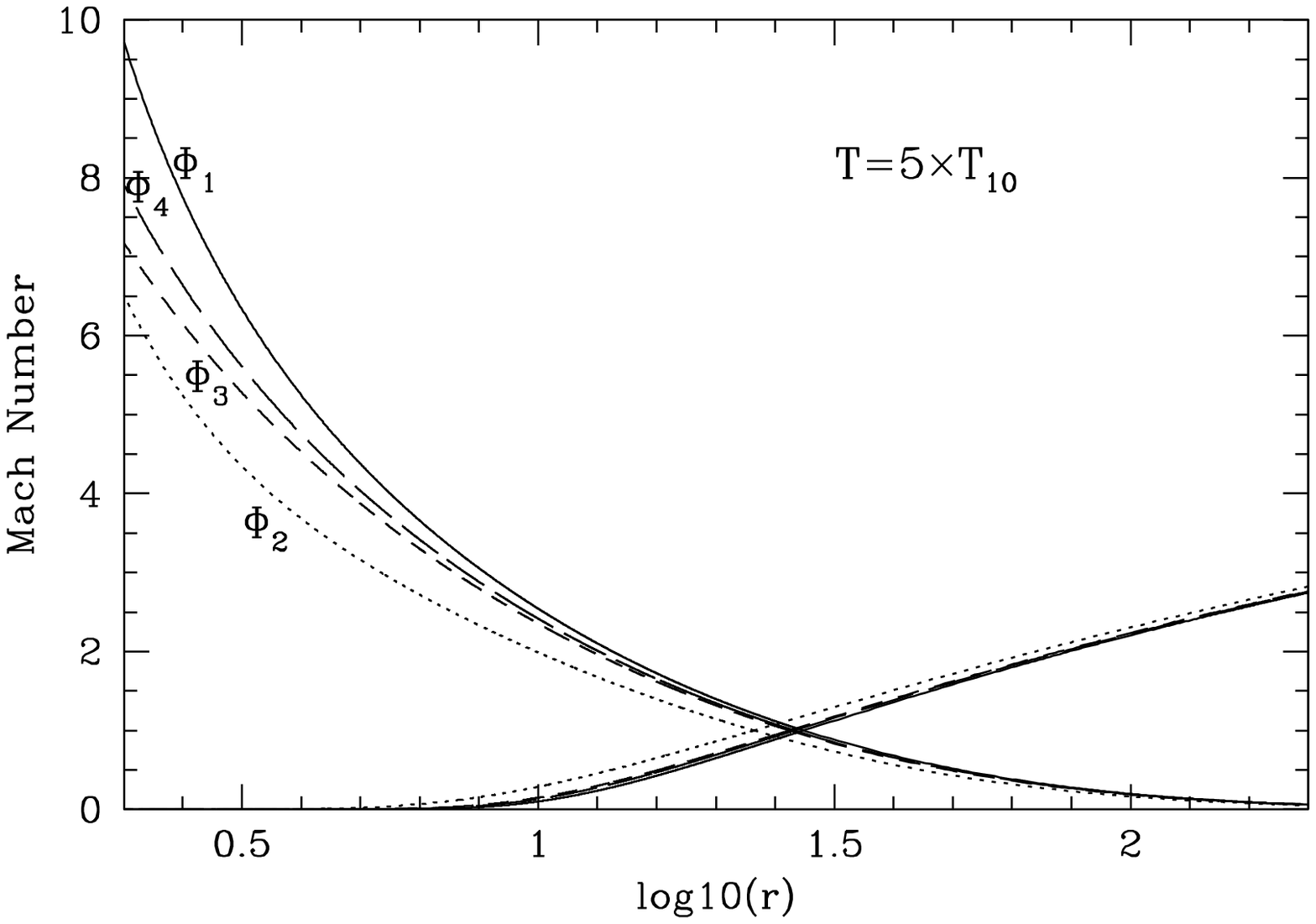,height=20cm,width=20cm}}}
\noindent {{\bf Fig. 2:}\\
{\bf Aveek Sarkar and Tapas K. Das\\
One Parameter Solution of Spherically Symmetric Accretion in Various
Pseudo-Schwarzschild Potentials}}
\end{figure}
\newpage
\begin{figure}
\vbox{
\vskip -5.1cm
\centerline{
\psfig{file=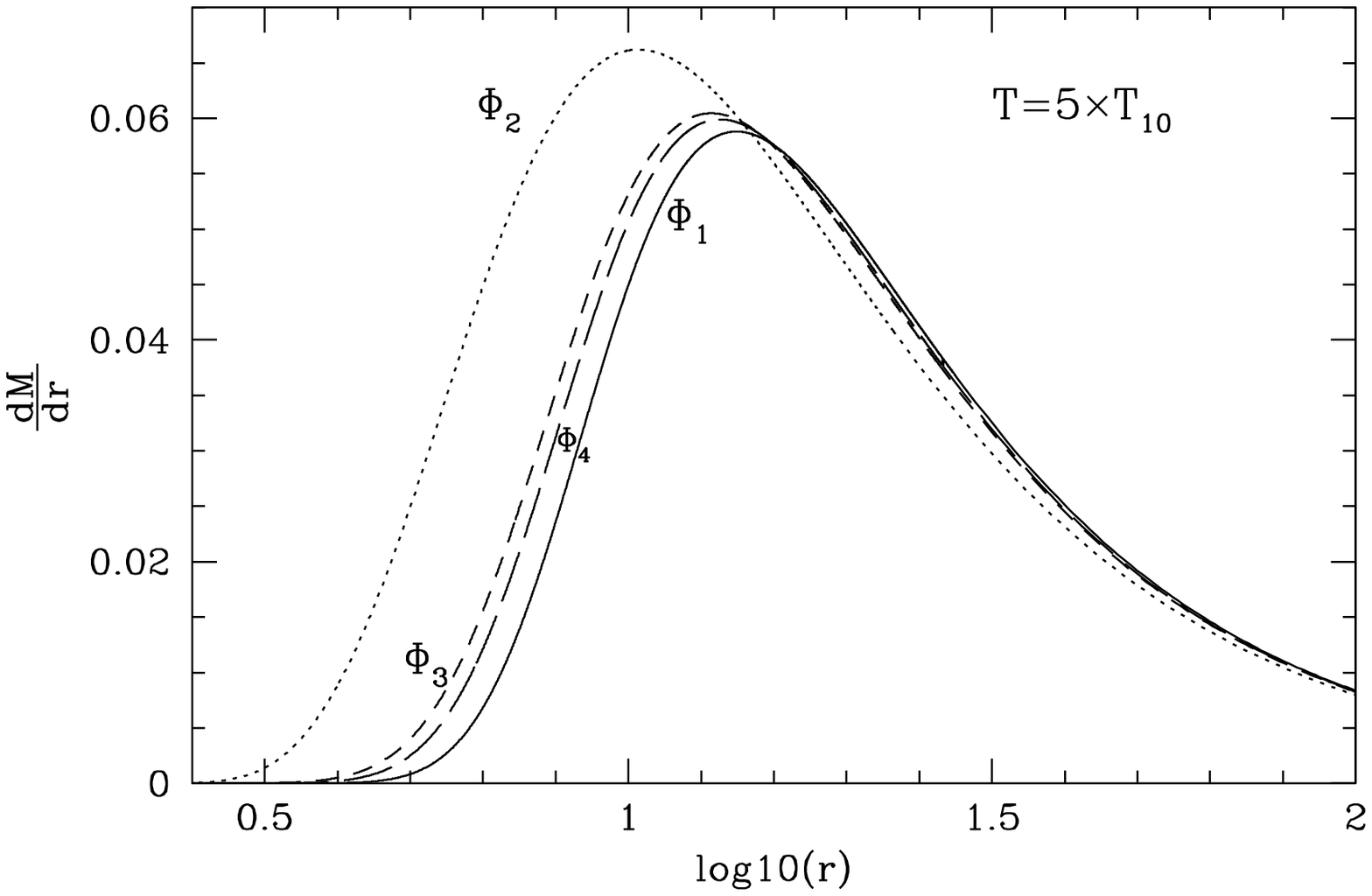,height=20cm,width=20cm}}}
\noindent {{\bf Fig. 3:}\\
{\bf Aveek Sarkar and Tapas K. Das\\
One Parameter Solution of Spherically Symmetric Accretion in Various
Pseudo-Schwarzschild Potentials}}
\end{figure}
\newpage
\begin{center}
{\large\bf Figure Captions}\\[1cm]
\end{center}
\begin{enumerate}
\item {\bf Fig. 1:} Variation of sonic point with flow temperature for various $\Phi_{i=1-4}$'s.
While flow temperature $T$ (in units of $T_{10}=10^{10}~{^o\!K}$) is plotted
(in logarithmic scale) along x axis, sonic point $r_c^i$ (in logarithmic
scale) is plotted along y axis.
\item {\bf Fig. 2:} 
The integral curves of motion drawn for various $\Phi_{i=1-4}$. While the radial
distance from the black hole (in units of $r_g$) is plotted along x axis
in logarithmic scale, Mach number $M$ of the flow is plotted along y axis.
The sonic points for various $\Phi_{i=1-4}$'s are obtained as
$\left\{r_c^1,r_c^2,r_c^3,r_c^4\right\}=\left\{28.05, 23.14, 26.6,
27.09\right\}r_g$ respectively.
\item {\bf Fig. 3:} 
Variation of gradient of Mach number for a given $T$ with respect to spatial
distance. While the radial distance from the accretor (in units of $r_g$) is
plotted along x axis in logarithmic scale, the spatial rate of change of
Mach number $\left(\frac{dM}{dr}\right)$ is plotted along y axis. It is clear
that the curves obtained for all $\Phi_{i=1-4}$ produces a `peak' location of
which is designated as $r_p$ in eqn. (8). $r_p$ is different for different
$\Phi_i$'s for a particular given flow temperature $T$.
\end{enumerate}

\end{document}